\documentclass[12pt]{article}
\usepackage{amsmath}
\usepackage{amstext,amssymb,amsfonts, graphicx,color,subfigure,slashed,bbold}
\usepackage[T1]{fontenc} 
\usepackage[sort]{cite}
\advance\voffset by -1.5cm
\advance\hoffset by -1.25cm
\newcommand{\sm}{\theta}

\definecolor{darkgreen}{rgb}{0,0.6,0}

\def\be{\begin{equation}}
\def\ee{\end{equation}}
\def\ba{\begin{eqnarray}}
\def\ea{\end{eqnarray}}

\newcommand{\ddd}{\displaystyle}

\newcommand{\vektor}{A}
\newcommand{\axial}{{A_5}}
\newcommand{\gluon}{G}
\newcommand{\dirac}{\slashed{D} - im}
\newcommand{\parv}{\alpha}
\newcommand{\para}{\epsilon}
\newcommand{\bra}[1]{\langle #1|}
\newcommand{\ket}[1]{|#1\rangle}

\newcommand{\comm}[2]{\left[#1, #2\right]}
\newcommand{\anticomm}[2]{\left\{#1, #2\right\}}
\newcommand{\C}{C_\gamma}
\newcommand{\Cg}{C_g}

\newcommand{\nn}{{\nonumber}}

\def\<{\langle}
\def\>{\rangle}

\usepackage{color}

\usepackage[english, USenglish]{babel}
\usepackage{epsfig}
\usepackage{slashed}

\begin{document}


\thispagestyle{empty}
\renewcommand{\thefootnote}{\fnsymbol{footnote}}
{\hfill \parbox{3.5cm}{

 DESY 12-126 \\
 ArXiv:1208.0012
}}

\bigskip\bigskip\bigskip

\begin{center} \noindent \Large \bf
Chiral superfluidity of the quark-gluon plasma
\end{center}

\bigskip\bigskip\bigskip\bigskip

\centerline{ \normalsize \bf
Tigran Kalaydzhyan\footnote[2]{\noindent \tt e-mail: tigran.kalaydzhyan@desy.de}
}

\bigskip\bigskip

\centerline{\it DESY Hamburg, Theory Group, Notkestrasse 85, 22607 Hamburg, Germany}
\centerline{\it ITEP Moscow, B. Cheremushkinskaya 25, 117218 Moscow, Russia}

\bigskip\bigskip\bigskip\bigskip\bigskip\bigskip

\renewcommand{\thefootnote}{\arabic{footnote}}

\centerline{\bf Abstract}
\medskip

{In this paper we argue that the strongly coupled quark-gluon plasma can be considered as a chiral superfluid. The ``normal'' component of the fluid is the thermalized matter in common sense, while the ``superfluid'' part consists of long wavelength (chiral) fermionic states moving independently. We use several nonperturbative techniques to demonstrate that. First, we analyze the fermionic spectrum in the deconfinement phase ($T_c < T \lesssim 2\,T_c$) using lattice (overlap) fermions and observe a gap between near-zero modes and the bulk of the spectrum. Second, we use the bosonization procedure with a finite cut-off and obtain a dynamical axion-like field out of the chiral fermionic modes. Third, we use relativistic hydrodynamics for macroscopic description of the effective theory obtained after the bosonization. Finally, solving the hydrodynamic equations in gradient expansion, we find that in the presence of external electromagnetic fields the motion of the ``superfluid'' component gives rise to the chiral magnetic, chiral electric and dipole wave effects. Latter two effects are specific for a two-component fluid, which provides us with crucial experimental tests of the model.}

\bigskip\bigskip


\setcounter{tocdepth}{2}
\flushbottom

\setcounter{equation}{0}
\section{Introduction}

Chiral properties of the strongly coupled quark-gluon plasma (sQGP) have attracted much attention in light of recent measurements performed within the heavy-ion programs at RHIC and LHC. The analysis of charge-dependent azimuthal correlations by STAR \cite{STAR,STAR1}, PHENIX \cite{PHENIX} and ALICE \cite{ALICE} collaborations suggests a possible local $\mathcal{P}$-violation in strong interactions, which manifests itself as an asymmetry in the charged particle production with respect to the reaction plane. One of the most popular theoretical approaches to study the observed phenomena is to seek for new electric currents of a specified direction in the QGP phase (see e.g. \cite{Kharzeev:2010gr,Warringa,Kharzeev:2007jp}). Since the physics of sQGP is essentially nonperturbative, there is a lack of models based on the first-principle calculations, i.e. starting from the QCD Lagrangian. In our paper we establish such a model for sQGP in the range of temperatures $T_c < T \lesssim 2\,T_c$, i.e. slightly above the deconfinement transition. We choose the range mainly because of two reasons: first, it is estimated to be typical for sQGP at RHIC \cite{Adare:2008ab}; and second, we can use hydrodynamic models to describe the system \cite{Shuryak_fluid,Shuryak_fluid1,RHICreport,RHICreport1,Shuryak_magcomp,Chernodub_magcomp,Chernodub_magcomp1}. 

\begin{figure}[t]
     \centering
     \subfigure[\label{below}]
     {\includegraphics[width=4cm]{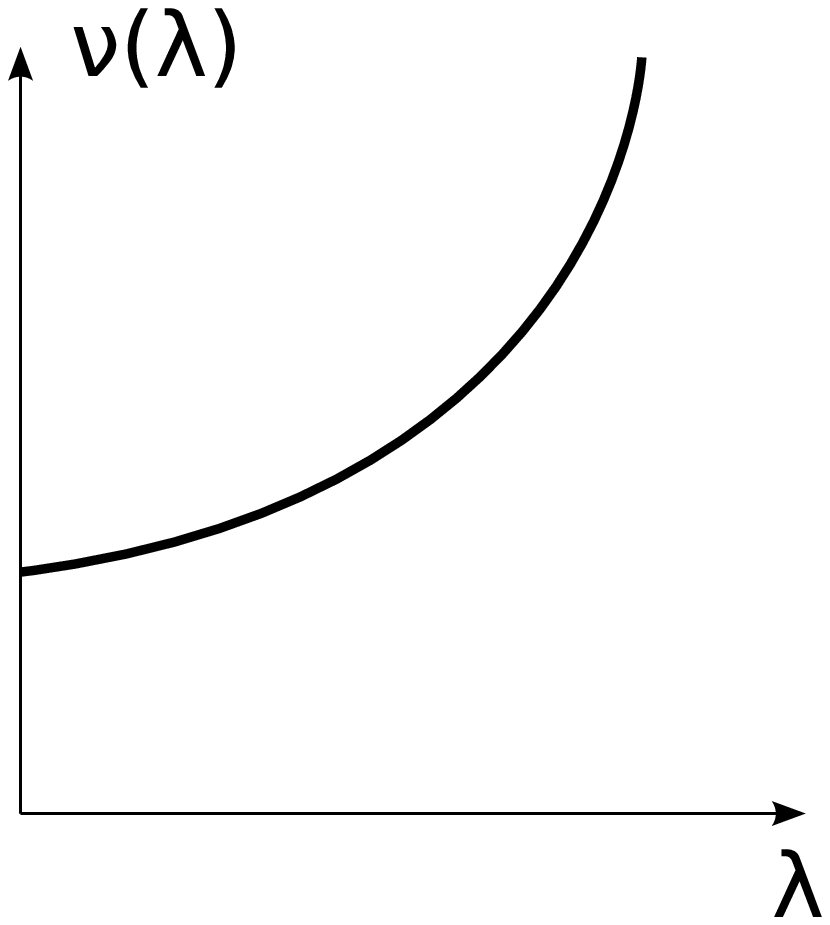}}\hspace{1cm}
     \centering
     \subfigure[\label{above}]
     {\includegraphics[width=4cm]{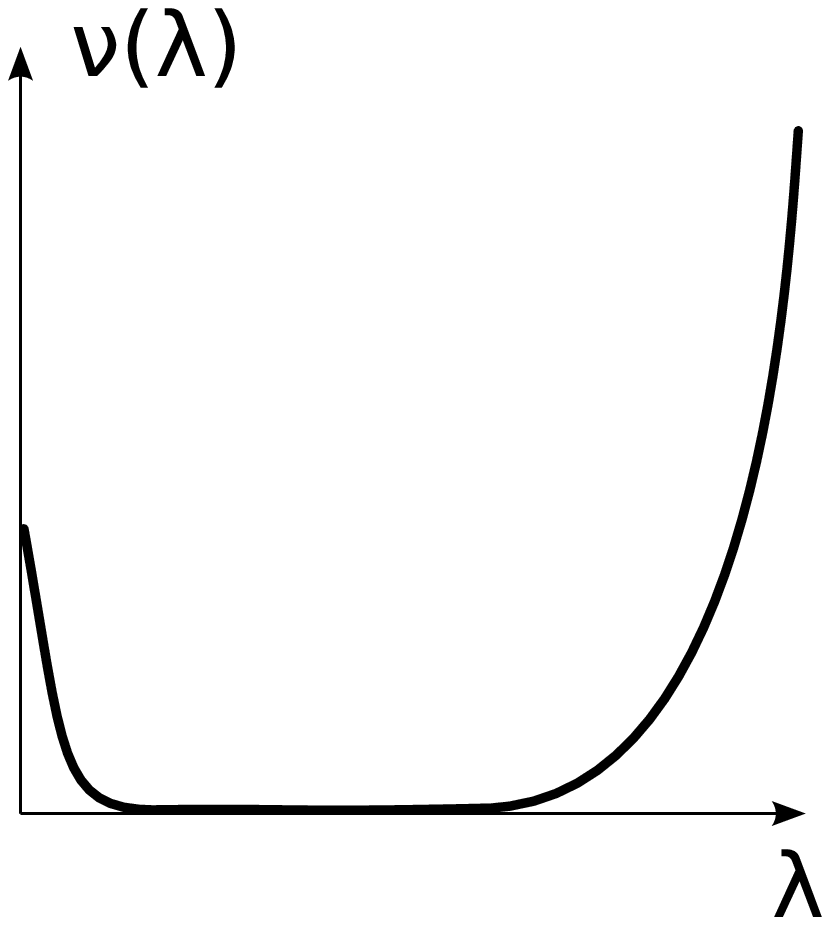}}\hspace{1cm}
     \centering
     \subfigure[\label{high}]
     {\includegraphics[width=4cm]{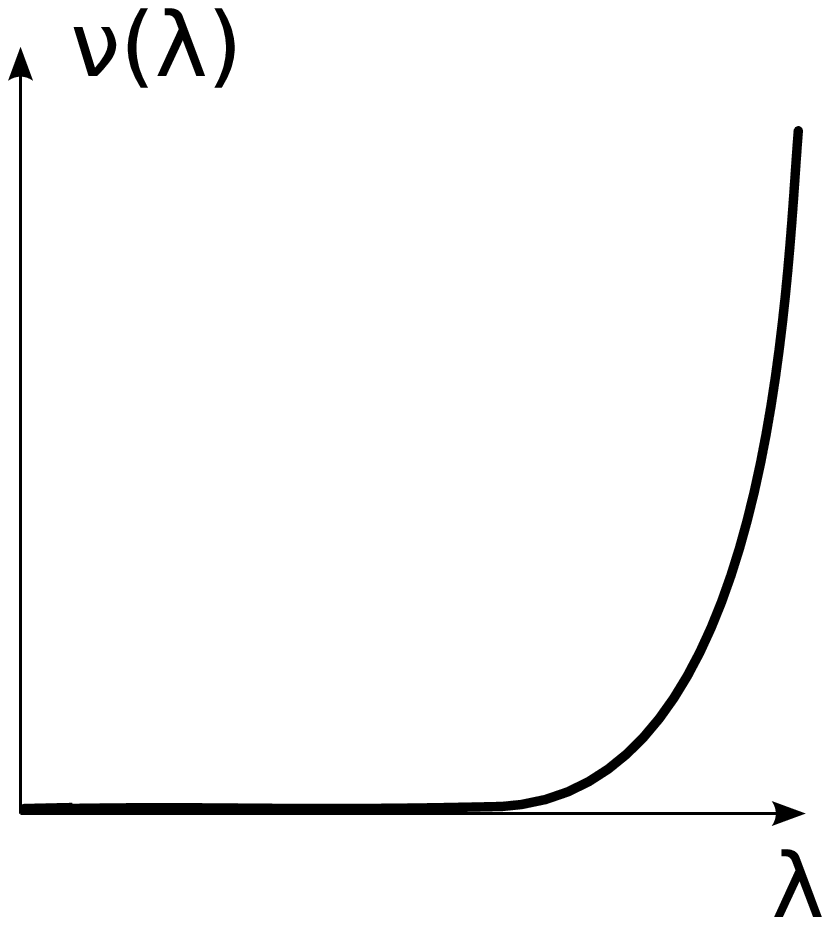}}
\caption{\label{spectrum} Fermionic spectrum of the chirally symmetric Dirac operator in a finite volume for $T<T_c$ (left), $T_c < T\lesssim 2\,T_c$ (center) and $T > 2\,T_c$ (right).}
\end{figure}

Lattice calculations \cite{Edwards,spectra2,spectra4} demonstrate that the Dirac spectrum for massless quarks at these temperatures contains a peak near zero virtuality separated by a gap from the bulk of the spectrum
\footnote{Calculations are performed in the quenched approximation. In fact, to make it more realistic, one needs a full QCD calculation with chemical potentials imitating initial conditions of the nuclear collisions} 
(Fig.~\ref{above} and Fig.~\ref{magspectrum}). Therefore, it is natural to introduce a two-component fluid model for the sQGP: one component, carrying chiral properties of the fluid, and the rest, corresponding to the bulk of the spectrum. In our case it turns out that the fluid is described by a system of equations similar to ones of a relativistic superfluid \cite{Son2000,Herzog:2008he}. This fact, of course, does not directly lead to the conventional superfluidity, since the ``normal'' and ``superfluid'' components in our case carry different $\mathrm{U(1)}$ charges. We also do not assume any hidden symmetry spontaneously broken in the system. Instead, we treat the term ``superfluidity'' in a phenomenological sense traced back to the Landau's formulation \cite{landau}, i.e. a combination of two independent (curl-free and ``normal'') motions of the fluid separated by an energy gap.

Before proceeding with a formal derivation let us provide a few hints supporting the existence of the ``superfluid'' component. In Refs. \cite{Horvath,Horvath1,Horvath2,Horvath3,Horvath4,Horvath5} it has been discovered and confirmed by \cite{Ilgenfritz:2007xu,Ilgenfritz:2007b} that topological charge density forms long-range coherent global structures around a locally one-dimensional network of strong fields (so-called ``skeleton''). Though the simulations were performed for low temperatures, the skeletons could very well survive slightly above the deconfinement transition, since there is a corresponding long-range order of gluonic fields \cite{Luscher} in the case of nonvanishing topological susceptibility. 
The extended character of the skeleton can be interpreted as a long-distance propagation of (quasi-) particles, carrying finite chirality and thus forming the chiral ``superfluid'' component.

In addition, the lattice data \cite{Horvath,Horvath1,Horvath2,Horvath3,Horvath4,Horvath5,Ilgenfritz:2007xu,Ilgenfritz:2007b,Kovalenko,fractal} suggest that the topological charge density itself (for uncooled configurations) is localized on low-dimensional defects with fractal dimension between 2 and 3, i.e. presumably on (percolating) central vortices (see also \cite{Aubin:2004mp,Gubarev:2005az} for a similar result based on the scalar density distribution and \cite{Reinhardt:2002cm,Hollwieser:2008tq} for a localization on vortex intersections). One can demonstrate \cite{Gattnar:2004gx,Gubarev:2005az,Hollwieser:2008tq,Bornyakov:2007fz} that removal of the central vortices eliminates all zero and chiral near-zero modes from the fermionic spectrum. The converse statement is also true: small eigenvalues can be generated at separate center vortex configurations \cite{Gattnar:2004gx}. It seems natural to consider the vortices as spatially one-dimensional ``guides'' for light fermions. Since massless fermions propagate in (1+1) dimensions with the same speed (speed of light), we can consider their bilinear combinations as bosonic excitations, even without assuming their interaction with each other. A similar effect of binding for light fermions along 1D defects, without making use of the Goldstone theorem, was considered in \cite{Tiktopoulos,Zhukovsky:2004ec,Chernodub:2012mu} for non-abelian fields and in \cite{CMW,Shovkovy} for a strong external magnetic field.

It is also known that monopole trajectories and central vortices populating QCD vacuum and leading to the confinement at low temperatures \cite{confinement} become Euclidean-time oriented (static) at high temperatures \cite{Engelhardt:1999fd}. Static nature of the vortices makes it possible to continue them to the Minkowski space-time by considering their intersections with planes of constant Euclidean time. These intersections are in general percolating strings, which, according to the polymer representation of the field theory \cite{polymer,polymer1}, can be interpreted as a 3D scalar field with non-vanishing vacuum expectation value. If this field is complex, then its phase can describe a new Goldstone mode in the deconfinement forming the superfluid component of sQGP \cite{Zakharov_twocomp,Zakharov_twocomp1}. Taking into account the vortex picture of the topological charge generation \cite{Engelhardt:2000wc,Engelhardt:2000wc1} it seems interesting to apply the arguments of \cite{Zakharov_twocomp,Zakharov_twocomp1} to the chiral superfluidity (we do not implement it here).

Other arguments in favor of the superfluidity were presented in recent analysis in the framework of stringy models \cite{Zakharov_superfluid,Zakharov_superfluid1,Zakharov_superfluid2}, where a new formulation of the superfluidity with vanishing chemical potential has been suggested.

In what follows we do not specify the way quarks are bound to each other and form a ``superfluid'' component, leaving the question open for the further studies (possible examples are given above and in the main text). Regardless of the nature of the ``superfluid'' component, we obtain the same universal phenomenological predictions, namely the chiral magnetic, chiral electric and dipole wave effects.

\section{Derivation of the effective Lagrangian}\label{derlag}

It is known that many of the essential properties of the QCD vacuum (such as the value of the chiral condensate, electric and magnetic characteristics, etc.)
can be determined from the IR part of the fermionic spectrum \cite{Banks:1979yr,Leutwyler,Buividovich:2009ih,Buividovich:2009wi,Buividovich:2009wi1,Ilgenfritz:2007xu}.
This makes it possible to introduce a finite cut-off $\Lambda$ for the fermionic spectrum, without affecting
the values of the observables (indeed, our phenomenological results do not depend on $\Lambda$, see Section~\ref{tests}).
In this section we perform the bosonization procedure with a finite
cut-off \cite{Andrianov} for the $\mathrm{SU(N_c)\times U_{em}(1)}$ theory and derive an effective Lagrangian. This procedure leads to appearance of
a dynamical axion-like field $\sm(x)$, which we identify with the propagating chirality (i.e. local difference between numbers of left- and right-handed fermions) in the system.
\subsection{The functional integral}
The gauge fields of the theory are represented by
\begin{align}
 \vektor_\mu = \vektor_\mu^0 T^0 + g\gluon_\mu^a T^a\equiv \vektor_\mu^{\hat{a}} T^{\hat{a}}, \qquad \axial_\mu = \axial_\mu^0 T^0\,,
\end{align}
where $\left\{T^a\, \left| \, a =\overline{1, (N_c^2-1)}\right.\right\}$ are the $SU(N_c)$ color matrices normalized by $\mathrm{Tr}(T^a T^b) = \delta^{a b}/2$ and
$T^0 = \mathbb{1}$. Here $\vektor_\mu^0$ and $\axial_\mu^0$ play a role of abelian fields, while $\gluon_\mu^a$ are gluonic fields.
The axial-vector field $\axial_\mu$ is an auxiliary external field and will be turned off at the end of the procedure.
The Euclidean functional integral for Dirac fermions ($N_f = 1$) in external vector $\vektor_\mu(x)$ and axial-vector $\axial_\mu(x)$ fields is given by
\begin{align}\label{Z}
 Z(\vektor, \axial) &= \int D \bar\psi D\psi\, \exp\left\{ -\int_V d^4x\, \bar\psi (\dirac) \psi \right\} = \det (\dirac)\,,\\
\slashed{D} &= -i(\slashed{\partial} + \slashed{\vektor} + \gamma_5\slashed{\axial})\,.\nonumber
\end{align}
The eigenvectors of $\slashed{D}$ are defined by\footnote{The Dirac operator $\slashed{D}$ should be Hermitian, therefore formally the axial field has to be rotated $\axial \rightarrow i \axial$ from the beginning and then rotated back $\axial \rightarrow -i \axial$ in the final result. We stick to conventions of \cite{Andrianov} and choose Hermitian gamma-matrices $\gamma^\dag = \gamma$ and $\gamma_5 \equiv - \gamma_0\gamma_1\gamma_2\gamma_3$, while in Fujikawa's earlier works \cite{Fujikawa,Fujikawa1} they are anti-Hermitian.}
\begin{align}
\slashed{D} \psi_n = \lambda_n \psi_n\,.
\end{align}
They form a complete orthonormal basis $\{\psi_n|n \in \mathbb{N}\}$ in the space of square-integrable spinors. According to
the general prescription \cite{Andrianov,Fujikawa,Fujikawa1}, we expand $\psi$ and $\bar\psi$ into this basis and cut the summation after first $N$ basis vectors:\footnote{We do not consider zero-modes, otherwise the functional integral (\ref{Z}) will vanish in the chiral limit. We are not interested in the \textit{global} topological properties of the system.}
\begin{align}
\psi(x) \rightarrow \sum\limits_{n=1}^N a_n\psi_n(x)\,,\qquad\qquad \bar\psi(x) \rightarrow \sum\limits_{n=1}^N \psi^{\dag}_n(x)b_n\,.
\end{align}
Then the functional integral (\ref{Z}) can be rewritten as
\begin{align}
& Z_N = \int \prod\limits_{n=1}^N d b_n\,d a_n\, \exp\left\{ -\sum\limits_{n,k=1}^N\, b_n \bra{\psi_n} \dirac \ket{\psi_k} a_k\right\} = \det (\dirac)_N\,,\\
&(\dirac)_N \equiv 1 - P_N + P_N (\dirac) P_N, \qquad P_N \equiv \sum\limits_{n=1}^N \ket{\psi_n}\bra{\psi_n}\,.\nonumber
\end{align}
From the gauge invariance of $(\dirac)_N$ it follows that the projector $P_N$ commutes with the Dirac operator $\left[ \slashed{D}, P_N\right]=0$, which will be used in the further calculations.

For the regularization we need to introduce a mass parameter $\Lambda$,
to be discussed below. It will be defined via $|\lambda_N|<\Lambda<|\lambda_{N+2}|$. The projector $P_N$ can then be replaced by
\begin{align}
 P_\Lambda \equiv \theta \left( 1 - \ddd\frac{\slashed{D}^2}{\Lambda^2} \right) = \int\limits_{-\infty}^{+\infty} \ddd\frac{d \zeta}{2\pi i(\zeta - i\varepsilon)} \exp\left[i\zeta (1 - \slashed{D}^2/\Lambda^2) \right]\,,\label{longprojector}
\end{align}
where $\theta(x)$ is the Heaviside step function. We also replace all the indices $N$ by $\Lambda$.

\subsection{Vector currents conservation}

The functional integral (\ref{Z}) is invariant under the gauge transformation
\begin{align}
\left\{
\begin{array}{l c l}
\vektor_\mu &\rightarrow& \vektor_\mu + \partial_\mu \parv + \comm{\vektor_\mu}{\parv}\,,\\
\axial_\mu &\rightarrow& \axial_\mu\,,\\
\psi &\rightarrow& (1 - \parv) \psi\,,\\
\bar\psi &\rightarrow& \bar\psi (1 + \parv)\,,
\end{array}
\right.\label{vector_tr}
\end{align}
where $\parv = \parv^{\hat{a}} T^{\hat{a}}$. Therefore,
\begin{align}
 Z_\Lambda(\vektor) = Z_\Lambda(\vektor + \partial \parv) \simeq Z_\Lambda(\vektor) + \int d^4 x\, \ddd\frac{\delta Z_\Lambda(\vektor)}{\delta \vektor^{\hat{a}}_\mu(x)} \cdot \partial_\mu \parv^{\hat{a}}(x)\,.
\end{align}
After integration by parts we get
\begin{align}
\partial_\mu\ddd\frac{\delta Z_\Lambda(\vektor)}{\delta \vektor^{\hat{a}}_\mu(x)} =0,
\end{align}
which means that the vector currents
\begin{align}
 j_{\Lambda}^{\mu\,{\hat{a}}}(x) \equiv - \ddd\frac{1}{Z_\Lambda}\frac{\delta Z_\Lambda(\vektor)}{\delta \vektor^{\hat{a}}_\mu(x)} = i\, \mathrm{Tr}\left(\gamma^\mu T^{\hat{a}} \bra{x} \ddd\frac{P_\Lambda}{(\dirac)_\Lambda} \ket{x} \right)
\end{align}
are conserved. To derive the last expression one can use the equivalences $\mathrm{exp}\,\mathrm{Tr}\,\mathrm{ln}(\cdot) = \mathrm{det}(\cdot)$ and
\begin{align}
\ddd\frac{1}{(\dirac)_\Lambda} = 1-P_\Lambda + \ddd\frac{P_\Lambda}{\dirac}\label{inversedirac}
\end{align}
as well as the general identity for projectors
\begin{align}
 P P' P = P (P^2)' P = 2 P P' P = 0\,,\label{projidentity}
\end{align}
where the prime denotes a derivative of $P$ with respect to an arbitrary parameter.

\subsection{Anomaly for the axial current}
Let us perform the chiral transformation of the functional integral
\begin{align}
\left\{
\begin{array}{l c l}
\vektor_\mu &\rightarrow& \vektor_\mu\,,\\
\axial^0_\mu &\rightarrow& \axial^0_\mu + \partial_\mu \para\,,\\
\psi &\rightarrow& (1 + \gamma_5\para) \psi\,,\\
\bar\psi &\rightarrow& \bar\psi (1 + \gamma_5\para)\,,
\end{array}
\right.\label{chiral_tr}
\end{align}
with $\para = \para^0 T^0$. Then in analogy with the previous section we get
\begin{align}
 Z_\Lambda(\axial) = Z_\Lambda(\axial + \partial \para, \para) \simeq Z_\Lambda(\axial) + \int d^4 x\, \ddd\frac{\delta Z_\Lambda(\axial, \para)}{\delta \axial^0_\mu(x)} \cdot \partial_\mu \para(x) + \int d^4 x\, \ddd\frac{\delta Z_\Lambda(\axial, \para)}{\delta \para(x)}\cdot \para(x)\,.\label{chiral_expand}
\end{align}
The second term here is due to a nontrivial transformation of the integration measure of the functional integral, and was absent in the case of vector current.
The projector $P_\Lambda$ and combined operator $\dirac$ also transform under the chiral rotations:
\begin{align}
P_\Lambda \rightarrow (1 + \gamma_5\para)P_\Lambda(1 + \gamma_5\para)\,, \qquad (\dirac)\rightarrow (1 + \gamma_5\para)(\dirac)(1 + \gamma_5\para)\,.\label{rotproj}\\
 \ddd\frac{\delta P_\Lambda(x)}{\delta \para(x')} = \anticomm{\gamma_5}{P_\Lambda(x)}\delta(x-x') \,, \qquad \ddd\frac{\delta (\dirac)(x)}{\delta \para(x')} = -2 i m \gamma_5\delta(x-x')\,.\label{rotdirac}
\end{align}
If we introduce the axial current as
\begin{align}
  j_{\Lambda}^{5 \mu}(x) \equiv - \ddd\frac{1}{Z_\Lambda}\frac{\delta Z_\Lambda(\axial)}{\delta \axial^0_\mu(x)}\,,
\end{align}
then from (\ref{chiral_expand}) it follows that
\begin{align}
\partial_\mu j_\Lambda^{5\,\mu} = -\ddd\frac{1}{Z_\Lambda}\ddd\frac{\delta Z_\Lambda(\axial)}{\delta \para(x)}\,.
\end{align}
The latter is equal to
\begin{align}
-\ddd\frac{1}{Z_\Lambda}\ddd\frac{\delta Z_\Lambda(\axial)}{\delta \para(x)} =& -\mathrm{Tr}\left[\ddd\frac{1}{(\dirac)_\Lambda}\left( -\ddd\frac{\delta P_\Lambda}{\delta\para}+ 2\ddd\frac{\delta P_\Lambda}{\delta\para}(\dirac)P_\Lambda
+ P_\Lambda \ddd\frac{\delta (\dirac)}{\delta\para} P_\Lambda \right)\right]\nonumber\\
 =&\, 2 i m \, \mathrm{Tr}\left( \gamma^5 T^0 \bra{x} \ddd\frac{P_\Lambda}{(\dirac)_\Lambda} \ket{x}\right) + 2\, \mathrm{Tr} (\gamma^5 \bra{x} P_\Lambda \ket{x})\,,
\end{align}
where we used (\ref{inversedirac}, \ref{projidentity}, \ref{rotproj}, \ref{rotdirac}) and the
commutation properties of $\slashed{D}$.
By means of (\ref{longprojector}), the second matrix element
in this formula can be expressed as
\begin{align}
&\mathrm{Tr} (\gamma^5 \bra{x} P_\Lambda \ket{x}) = \mathrm{Tr}\left\{\gamma^5 \int\ddd\frac{d^4 k}{(2\pi)^4} \int\limits_{-\infty}^{+\infty} \ddd\frac{d \zeta}{2\pi i(\zeta - i\varepsilon)} \mathrm{e}^{-ikx} \mathrm{e}^{i\zeta (1 - \slashed{D}^2/\Lambda^2)} \mathrm{e}^{ikx} \right\} \\
&~~~= \mathrm{Tr}\left\{\gamma^5 \int\ddd\frac{d^4 k}{(2\pi)^4} \int\limits_{-\infty}^{+\infty} \ddd\frac{d \zeta}{2\pi i(\zeta - i\varepsilon)} \exp\left[i\zeta \left(1 - \ddd\frac{k^2}{\Lambda^2} - \frac{2}{\Lambda^2}k^\mu D_\mu(x) - \frac{\slashed{D}^2}{\Lambda^2}\right)\right] \right\}\\
&~~~= -\int\ddd\frac{d^4 \tilde k}{(2\pi)^4}\int\limits_{-\infty}^{+\infty} \ddd\frac{\zeta^2 d \zeta}{4\pi i(\zeta - i\varepsilon)}\mathrm{e}^{i\zeta(1-\tilde k^2)} \mathrm{Tr}\left[\gamma_5 ( \slashed{D}^4 + 4\Lambda^2\, (\tilde k^\mu D_\mu)^2 + \Lambda^2\frac{2i}{\zeta} \slashed{D}^2)\right] \nn\\&~~~~~~+ O\left(\ddd\frac{1}{\Lambda^2}\right)\,,
\end{align}
where $\tilde k \equiv k/\Lambda$. Using the algebra of the gamma-matrices,
we write
\begin{align}
\slashed{D}^2 = -D_V^2 - \frac{1}{2}\gamma^{[\mu}\gamma^{\nu]} \vektor_{\mu\nu} +\axial_\mu^2 - (\partial^\mu \axial_\mu) \gamma^5 - \gamma^{[\mu}\gamma^{\nu]} \left( D_{V \mu}\axial_\nu - \axial_\mu D_{V \nu} \right)\gamma^5 \,,
\end{align}
where $D_V \equiv -i (\partial + \vektor)$, and finally obtain
\begin{align}
& \partial_\mu j_{\Lambda}^{5 \mu} = \,2m \rho^5_\Lambda +\frac{1}{16\pi^2} \varepsilon^{\mu\nu\lambda\kappa} \mathrm{Tr}\left( \vektor_{\mu\nu}\vektor_{\lambda\kappa} + \frac{1}{3} \axial_{\mu\nu}\axial_{\lambda\kappa} \right)\nonumber\\
& + \frac{1}{4\pi^2}\mathrm{Tr}\left( \partial_\mu \partial^\mu \partial^\nu \axial_\nu + \frac{2}{3} \anticomm{\anticomm{\partial^\mu\axial^\nu}{\axial_\nu}}{\axial_\mu} + \frac{1}{3} \anticomm{\partial^{\mu}\axial_\mu}{\axial^\nu \axial_\nu} + \frac{2}{3}\axial^\mu\partial^\nu\axial_\nu\axial_\mu \right)\nonumber\\
& + \ddd\frac{\Lambda^2}{2\pi^2} \mathrm{Tr}\left(\partial^\mu \axial_\mu\right)  + O\left(\ddd\frac{1}{\Lambda^2}\right)\,,\label{j5anomaly}
\end{align}
where the chirality $\rho^5_\Lambda$ is defined by
\begin{align}
 \rho^5_\Lambda \equiv i\, \mathrm{Tr}\left(\gamma^5 T^0\bra{x} \ddd\frac{P_\Lambda}{(\dirac)_\Lambda} \ket{x}\right)
\end{align}
and the field strengths are
\begin{align}
\vektor_{\mu\nu} &= \partial_\mu \vektor_\nu - \partial_\nu \vektor_\mu + \comm{\vektor_\mu}{\vektor_\nu},\nonumber\\
\axial_{\mu\nu} &= \partial_\mu \axial_\nu - \partial_\nu \axial_\mu + \comm{\vektor_\mu}{\axial_\nu} - \comm{\vektor_\nu}{\axial_\mu} = \partial_\mu \axial_\nu - \partial_\nu \axial_\mu.\nonumber
\end{align}
The first line in (\ref{j5anomaly}) is the usual chiral anomaly with a mass-dependent term and the topological term. The rest of (\ref{j5anomaly}) can be subtracted away by adding to the Lagrangian the following gauge-invariant term:
\begin{align}
\Delta\mathcal{L}  = \frac{1}{12\pi^2}\mathrm{Tr}\left(3\Lambda^2\,(\axial_\mu \axial^\mu)
+ \frac{1}{2} \left(\partial_\mu \axial^\mu\right)^2 -\left(\axial_\mu \axial^\mu\right)^2\right)\label{massterm}\,.
\end{align}

\subsection{Axionic Lagrangian}

The external axial-vector field $\axial_\mu$ should be now switched off, which is equivalent to considering the pure gauge $\axial^0_\mu = \partial_\mu \sm$, since one can always generate a pseudoscalar field $\sm$ by the \textit{local} chiral rotation $\slashed{\vektor} \rightarrow \slashed{\vektor} + \gamma^5\slashed{\partial}\sm$. In other words, the quadratic term in (\ref{massterm}) is required by having the chiral transformation consistent with the correct form of the chiral anomaly. 
In the chiral limit $m \rightarrow 0$ the total effective Euclidean Lagrangian is given then by\footnote{Currents $j^{\hat a \mu}$ can be sourced by fermions from the scales above $\Lambda$.}
\begin{align}
{\cal L}^{(4)}_E = &\, \Delta\mathcal{L} +\frac{1}{4}G^{a\mu\nu}G_{\mu\nu}^{a} + \frac{1}{4}F^{\mu\nu}F_{\mu\nu} -j^{\hat a\,\mu} \vektor_\mu^{\hat a} - j^{5 \mu} \axial^0_\mu\nn\\
  = &\,  \Delta\mathcal{L} +\frac{1}{4}G^{a\mu\nu}G_{\mu\nu}^{a} + \frac{1}{4}F^{\mu\nu}F_{\mu\nu} -j^{0 \mu} \vektor^0_\mu - g j^{a\,\mu}\gluon^a_\mu -j_{\Lambda}^{5 \mu} \partial_\mu\sm \nn\\
  \sim &\,  \frac{1}{4}G^{a\mu\nu}G_{\mu\nu}^{a} + \frac{1}{4}F^{\mu\nu}F_{\mu\nu}
-j^{0 \mu} \vektor^0_\mu - g j^{a\,\mu}\gluon^a_\mu \nn\\
 & + \ddd\frac{\Lambda^2 N_c}{4\pi^2} \partial^\mu\sm\partial_\mu\sm\ +\frac{g^{2}}{16\pi^{2}}\sm\gluon^{a\mu\nu}\widetilde{\gluon}{}_{\mu\nu}^{a} + \frac{N_c}{8\pi^{2}}\sm F^{\mu\nu}\widetilde{F}{}_{\mu\nu} \nn\\
 & +\frac{N_c}{24\pi^2}\sm \Box^2 \sm - \frac{N_c}{12\pi^2} \left( \partial^\mu\sm\partial_\mu\sm\right)^2 \,,\label{quartic_L}
\end{align}
where we used the anomaly expression (\ref{j5anomaly}) and integrated by parts. Kinetic terms for the electromagnetic and gluonic fields could be introduced already in (\ref{Z}) and do not affect the derivation. As we see, the kinetic term for $\theta$ even being absent in (\ref{Z}) is generated dynamically
\footnote{The $\theta$-field does not appear in the
integration measure and hence should be treated as a classical low-energy excitation. It is possible due to the fact that the loop corrections by $\theta$-fields are finite in 4D and depend on the
powers of external momenta divided by $\Lambda$, i.e. such loop amplitudes are suppressed by powers of $\Lambda$.}, see also \cite{Rabinovici, Polyakov:1981rd}
 for similar examples. If we drop quartic terms and  replace conventions $\vektor^0_\mu \rightarrow \vektor_\mu$, $j^{0 \mu} \rightarrow j^\mu$, then the effective Lagrangian is reduced to
\begin{align}
{\cal L}^{(2)}_E =&\, \frac{1}{4}G^{a\mu\nu}G_{\mu\nu}^{a} + \frac{1}{4}F^{\mu\nu}F_{\mu\nu}
-j^\mu \vektor_\mu
 - g j^{a\,\mu}\gluon^a_\mu
\nonumber\\
&+ \ddd\frac{\Lambda^2 N_c}{4\pi^2} \partial^\mu\sm\partial_\mu\sm\ +\frac{g^{2}}{16\pi^{2}}\sm\gluon^{a\mu\nu}\widetilde{\gluon}{}_{\mu\nu}^{a} + \frac{N_c}{8\pi^{2}}\sm F^{\mu\nu}\widetilde{F}{}_{\mu\nu}\,.\label{Leucl}
\end{align}

This Lagrangian describes a generalization of the axion electrodynamics \cite{Wilczek:1987mv}, where the new terms are due to gluonic fields.
It seems interesting that the axionic field $\theta(x,t)$ appears within QCD coupled to QED, \textit{without} any further assumptions as e.g. the Peccei-Quinn mechanism \cite{PQ}.
Similarly to the `true' axion, $\theta(x,t)$ is a propagating dynamical field. However, the value of the decay constant $f = \ddd\frac{2\Lambda}{\pi}\sqrt{N_c}$ turns out to be
of order of scales appearing in QCD (see below), while in cosmological scenarios this value is usually around $10^9 - 10^{12}$ GeV. The formal similarity of (\ref{Leucl}) to the axion Lagrangian allows us
to derive an explicit expression for the mass of $\theta(x,t)$ \cite{McLerran:1990de,Witten:1979vv,Wantz:2009it},
\begin{align}
m_\theta^2 f^2 = {\chi(T)} \,,
\end{align}
where $\chi= \lim_{V \rightarrow \infty} \frac{\langle Q^2 \rangle}{V}$
is the topological susceptibility related to fluctuations of the topological
charge $Q$. Lattice simulations demonstrate that $\chi$
goes (almost) to zero at temperatures above the deconfinement
transition \cite{Alles,Edwards,spectra2,Hegde:2011zg,Hegde:2011zg1}, this behaviour is also confirmed within the interacting instanton liquid model \cite{Wantz:2009it}. Meson masses in the deconfinement interpolate between their values at $T \lesssim T_c$ and approximately twice the lowest quark Matsubara frequency (i.e.~$2\pi T$). These two facts allow us to consider the axion-like field $\theta(x,t)$  as a nearly-massless field, an essential requirement for a superfluid mode.

Instead of considering $\sm$ as a real particle we rather tend to interpret it as a collective excitation (quasi-particle) of the medium in the nonperturbative regime of QCD (e.g. combinations of chiral quarks).
As will be shown in next sections, the excitation carries chirality and can be considered as a 4D generalization of the Chiral Magnetic Wave \cite{CMW}.
These excitations can also be exactly massless (compare with phonons or sound waves) and, at the same time, not necessarily consisting of massless quantum particles,
appearing as Goldstone bosons of some broken (e.g. Peccei-Quinn-like) symmetry\footnote{Even if we assume that $\mathrm{U(1)_A}$ symmetry is
broken spontaneously, as suggested in \cite{Kapusta}, then the bosonization procedure does not describe $\eta'$, since
 it is a spin-0 particle, which can not carry chirality as $\theta$ does.}. A straightforward derivation of this collective solution
made out of quarks and gluons is not worked out at the moment. However, there are some evidences that it could exist, see e.g. comments on 
binary bound states in sQGP \cite{ShuryakZahed,ShuryakZahed1}, lattice results on a screened attractive force in the color-singlet channel \cite{screening} 
and ideas mentioned already in the Introduction.

\subsection{Interpretation of $\Lambda$}\label{interpretation}

The scale $\Lambda$ can be studied by considering $N_f = N_c = 1$ in the limit of a constant background $\mu=const$, $\mu_5=const$ and negligible anomaly (i.e. slow varying $\sm = i\mu_5 t_E$).
In this case, keeping also the quartic terms in (\ref{quartic_L}) for generality, we get
\begin{align}
\rho_5 = -\ddd\lim\limits_{t_E \rightarrow 0}\frac{\delta {\cal L}^{(4)}_E}{\delta\mu_5} = \frac{1}{2}\left(\frac{\Lambda}{\pi} \right)^2 \mu_5 + \frac{1}{3\pi^2}\mu_5^3\,. \label{our_rho}
\end{align}
In other words, the value of $\Lambda$ can be read off from the dependence of chirality $\rho_5$ on the chiral chemical potential $\mu_5$.
We consider here three existing examples one can find in the literature.
\begin{itemize}
 \item[(1)] At high temperatures, neglecting effects of gluons and assuming equilibrium, one can define the thermodynamic grand potential as \cite{Warringa}
\begin{align}
\Omega = \sum\limits_{s=\pm}\int\frac{d^3 p}{(2\pi)^3}\left[ \omega_{p,s} + T\sum_{\pm}\log (1 + e^{-\frac{\omega_{p,s}\pm \mu}{T}}) \right]\label{omega_potential}\,,
\end{align}
where $\omega_{p,s}^2 = (p+s\mu_5)^2 + m^2$ and one also assumes an approximate conservation of the axial charge.
Differentiating the grand potential with respect to $\mu_5$ and taking the massless limit one obtains \cite{Warringa}
\begin{align}
\rho_5 = \frac{1}{3}\left(T^2 + \ddd\frac{\mu^2}{\pi^2} \right)\mu_5 + \frac{1}{3\pi^2}\mu_5^3\,.\label{Kharzeev_rho}
\end{align}
Comparing (\ref{Kharzeev_rho}) with (\ref{our_rho}) we conclude
\begin{align}
 \Lambda = \pi\sqrt{\frac{2}{3}}\sqrt{T^2 + \frac{\mu^2}{\pi^2}}\qquad (\mbox{free quarks})\,.
\end{align}
In \cite{Kharzeev:2007jp,Warringa} this scale is compared with the inverse radius of a typical sphaleron at given temperature.
Notice, that in the high temperature limit $T\gg\mu$ and $T\gg\mu_5$ we get simplifications $\Lambda \propto T$ and $\rho_5 \propto \Lambda^2 \mu_5 \propto T^2 \mu_5$.

\item[(2)] In case of a strong external magnetic field ($eB >\mu_5^2/2$) one can construct the grand potential for fermions on the lowest Landau level \cite{Warringa}
\begin{align}
\Omega = \frac{eB}{4\pi^2}\int\limits_{-\infty}^{\infty} d^3 p_{||}\left[ \omega_{p} + T\sum_{\pm}\log (1 + e^{-\frac{\omega_{p}\pm \mu}{T}}) \right]\label{omega_potential1}\,,
\end{align}
where $\omega_{p}^2 = (p_{||} + \mu_5)^2 + m^2$ and $p_{||}$ denotes a component of momentum parallel to the magnetic field. This gives us $\ddd\rho_5 = \frac{eB}{2\pi^2}\mu_5$ and hence
\begin{align}
 \ddd\Lambda = 2\sqrt{eB}\qquad (\mbox{free quarks and strong }B)\,.
\end{align}
Upon the redefinition $\ddd\sm \rightarrow \frac{\pi}{\sqrt{2 N_c eB}}\sm$ the kinetic term for $\sm$ in the effective Lagrangian (\ref{quartic_L}) takes a canonical form $\ddd\frac{1}{2}(\partial_\mu\sm)^2$, while the quartic terms are suppressed by factors $1/B$ and $1/{B^2}$, respectively. 
Therefore, the bosonization procedure becomes exact in the limit $B\rightarrow\infty$ as in \cite{CMW}.

\item[(3)] To include effects of gluons one needs to perform a lattice calculation with finite $\mu_5$, which has been done in \cite{Yamamoto,Yamamoto1}. Slope of the curve $\rho_5=\rho_5(\mu_5)$ obtained in the paper\footnote{There is no \textit{a priori} introduced UV-cutoff in the paper since the inversion of the Dirac operator is done by means of the BiCGstab solver} is approximately one (in lattice units), which being translated to physical units and compared with (\ref{our_rho}) gives us
\begin{align}
  \ddd\Lambda \simeq 3\, \mathrm{GeV}\qquad (\mbox{dynamical lattice fermions}, N_f=2, N_c=3)\,.\label{yamamoto_formula}
\end{align}
Appearance of this scale (much larger than $\Lambda_{QCD}$) is not surprising, see e.g. \cite{Zakharov:2002md,Shuryak:1999qs}.
\end{itemize}
It is worth to mention that all three predictions even if being affected by either rought initial assumptions or lattice artifacts, still provide a \textit{finite} and reasonable value of $\Lambda$.

\subsection{Fermionic spectrum and chirality\label{superfluidity}}
The Dirac spectrum for massless fermions is schematically shown in Fig.~\ref{spectrum}.\footnote{We show here only the non-negative part of the spectrum. In the chiral limit it is symmetric with respect to the reflection $\lambda \rightarrow -\lambda$.} Lattice studies can be found in
 \cite{Edwards,spectra1,spectra2,Gubarev:2005az,spectra4}. Below
the critical temperature $T_c$ the spectrum is a continuously
growing function in $\lambda$.

\begin{figure}[t]
\centering
\includegraphics[angle=-90, width=8cm]{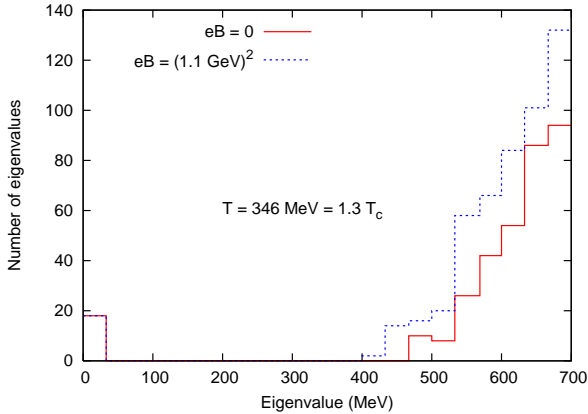}
\caption{\label{magspectrum} Typical fermionic spectrum in the deconfined phase as seen by a SU(3) quenched lattice simulation with tadpole-improved L\"uscher-Weisz action and overlap fermions ($\beta=8.45$, $a = 0.095\,\mathrm{fm}$, $V = 16^3\times 6$, $N_f=1, T_c \sim 260 MeV$).}
\end{figure}

In contrast, at $T_c<T\lesssim 2\,T_c$ the spectrum consists of three parts:
exact zero modes (and near-zero modes), followed by
a gap for the low-lying modes and a continuous spectrum starting from a finite $\lambda_B$. Presence of the near-zero peak can be interpreted as a manifestation of a (small) remaining chiral condensate \cite{Edwards,spectra4}, which is, howewer, not yet rigorously proven. Both the chiral condensate and topological susceptibility $\chi(T)$ are defined from the near-zero modes, since the exact zero modes do not survive in the thermodynamic limit \cite{Leutwyler,Hegde:2011zg,Hegde:2011zg1}. It is important, that $\chi(T)$ is small enough (to keep $m_\sm$ small), but still not zero (otherwise $\sm$ itself does not exist). This forces us to choose the window of temperatures $T_c<T\lesssim 2\,T_c$. At higer temperatures $T \gtrsim 2\,T_c$ the peak disappears completely and all the correlations between quarks supposed to be washed out by thermal effects. The gap width seems to be temperature dependent and grows with the temperature \cite{spectra2}. The right bound $\lambda_B$ is natural to identify with an effective quark mass, since on the corresponding fermionic mode
\begin{align}
\lambda_B^2 \psi_B = \slashed{D}^2 \psi_B = \slashed{p}^2 \psi_B = m_{eff}^2 \psi_B\,.
\end{align}
A very strong external magnetic field can slightly shift the right bound of the spectrum to the left\footnote{We are grateful to Victor Braguta for making this observation from our data} (see Fig.~\ref{magspectrum}), but for magnetic field strengths occuring in heavy ion collisions \cite{Kharzeev:2007jp} the principal shape of the spectrum remains
the same.

Let us demonstrate that chirality is determined by the first part of the spectrum. 
\begin{align}
 \rho^5_\Lambda &= \, i\, \mathrm{Tr}\left(\gamma^5\bra{x} \ddd\frac{P_\Lambda}{(\dirac)_\Lambda} \ket{x}\right) = i\sum\limits_{0 < |\lambda| < \Lambda}\frac{{\psi_\lambda}^\dagger \gamma^5 \psi_\lambda}{\lambda - im}\nonumber\\
&= i\sum\limits_{0 < \lambda < \Lambda}\frac{{\psi_\lambda}^\dagger \gamma^5 \psi_\lambda}{\lambda - im} + i\sum\limits_{0 < \lambda < \Lambda}\frac{{\psi_\lambda}^\dagger \gamma^5\gamma^5\gamma^5 \psi_\lambda}{-\lambda - im} = -2m \sum\limits_{0 < \lambda < \Lambda}\frac{{\psi_\lambda}^\dagger \gamma^5 \psi_\lambda}{\lambda^2 + m^2}\nonumber\\
&= -\int\limits_0^\Lambda d\lambda\, \nu(\lambda)\frac{2m}{\lambda^2 + m^2}{\psi_\lambda}^\dagger \gamma^5 \psi_\lambda\,,
\end{align}
where $\nu(\lambda)$ denotes the spectral density for an eigenvalue $\lambda$. Here, as before, we dropped the exact zero-modes.
Then, using the identity
\begin{align}
 \lim\limits_{m\rightarrow 0} \frac{m}{\lambda^2 + m^2} = \pi\delta(\lambda)
\end{align}
we find the following expression for the chirality
\begin{align}
 \lim\limits_{m\rightarrow 0} \rho^5_\Lambda = -2\pi \int\limits_0^\Lambda d\lambda\, \nu(\lambda) \delta(\lambda) {\psi_\lambda}^\dagger \gamma^5 \psi_\lambda =
-\pi \lim\limits_{\lambda\rightarrow 0} \nu(\lambda)\, {\psi_\lambda}^\dagger \gamma^5 \psi_\lambda\,.
\end{align}
From this expression we see that the chirality can be determined exclusively from the near-zero fermionic modes. It is suggested and tested on a lattice \cite{Edwards} that this part of the spectrum is originated from the zero modes of separate topological defects populating the vacuum, because interactions between the original zero modes break their degeneracy. So the exclusion of the \textit{exact} zero modes from our analysis does not affect much the main results (for additional arguments see also \cite{Buividovich:2009wi}).

\section{Quark-gluon plasma as a two-component fluid}

We now consider Minkowski version of the effective Lagrangian (\ref{Leucl}) for the quark-gluon plasma with one quark flavor\footnote{The metric we use has the signature $(- + + +)$},
\begin{align}
\label{Leff}
{\cal L}^{(2)} =& -\frac{1}{4}G^{a\mu\nu}G_{\mu\nu}^{a} - \frac{1}{4}F^{\mu\nu}F_{\mu\nu} -j^\mu \vektor_\mu  - g j^{a\,\mu}\gluon^a_\mu \nonumber\\
&- \ddd\frac{f^2}{2} \partial^\mu\sm\partial_\mu\sm\ +\frac{\Cg}{4}\sm\gluon^{a\mu\nu}\widetilde{\gluon}{}_{\mu\nu}^{a} + \frac{\C}{4}\sm F^{\mu\nu}\widetilde{F}{}_{\mu\nu}\,,
\end{align}
where, again, the decay constant is defined by $f = \ddd\frac{2\Lambda}{\pi}\sqrt{N_c}$ and the anomaly coefficients are given by $\C=\ddd\frac{N_c}{2\pi^{2}}$ and $\Cg=\ddd\frac{g^2}{4\pi^{2}}$.

Varying the Lagrangian (\ref{Leff}) with respect to $\theta$ and vector fields
$A_\mu$ and $G_\mu^a$, we obtain the following equations of motion:
\begin{align}
\partial^{\mu}\partial_{\mu} \theta &= -
 \frac{\C}{4 f^2} F^{\mu\nu}\widetilde{F}_{\mu\nu} - \frac{\Cg}{4 f^2} G^{\mu\nu a} \widetilde{G}^a_{\mu\nu} \,,\label{wavetheta}\\
\partial_{\mu}F^{\mu\nu} &= -j^{\nu}+\C(\partial_{\kappa}\theta)\tilde{F}^{\kappa\nu} \label{mcs-eom}\,,\\
\partial_{\mu}G^{\mu\nu a} &+ g f^{abc} G_\mu^b G^{\mu\nu c}  = -g j^{\nu a} + \Cg(\partial_{\kappa}\theta)\widetilde{G}^{\kappa\nu a} \label{grav-eom}\,,
\end{align}
where $f^{abc}$ are the structure constants of $SU(N)$. Bianchi identities are given by
\begin{align}
\partial_{\mu}\widetilde{F}^{\mu\nu} & = 0 \label{Bianchi}\,,\\
\partial_{\mu}\widetilde{G}^{\mu\nu a} + g f^{abc} G_\mu^b \widetilde{G}^{\mu\nu c} & = 0 \label{Bianchi2}\,.
\end{align}
Interestingly, $\theta(\vec x,t)$ obeys a wave equation which is sourced
by the $U(1)_A$ chiral anomaly. This makes it a four-dimensional generalization of the
so-called ``Chiral Magnetic Wave'' which has been recently proposed in \cite{CMW}.

\subsection{Hydrodynamic equations}\label{sechydro}

Here we follow a general idea of the hydrodynamic description of the anomalous phenomena in the quark-gluon plasma (read e.g. \cite{Zakharov:2012vv} for a review).
The hydrodynamic equations may now be derived from the
effective Lagrangian (\ref{Leff}) and the corresponding equations of motion (\ref{wavetheta})--(\ref{Bianchi2}).
Taking the divergence of (\ref{mcs-eom}) and using the Bianchi identity (\ref{Bianchi}), we obtain the conservation law for the electromagnetic current, $\partial_\nu j^\nu = 0$,
provided that the topology of the $\theta$-field is non-singular
with\footnote{delta-function on the r.h.s. of (\ref{trivialsol}) would generate an axionic string \cite{Callan}. For the relevant phenomenological consequences see \cite{Metlitski,Kirilin:2012mw}.
}
\begin{align} \label{trivialsol}
[\partial_\mu,\partial_\nu]\theta =0 \,.
\end{align}
Varying ${\cal L}^{(2)}$ with respect to
the axial-vector $-\partial_\mu \theta$, we obtain the axial current
\begin{align}
j^\mu_5 = f^2 \partial^\mu \theta \,.\label{js2}
\end{align}
This current satisfies the anomaly equation
\begin{align}
\partial_\mu j_5^\mu &= -\frac{\C}{4}
  F^{\mu\nu} \widetilde F_{\mu\nu} -
 \frac{\Cg}{4} G^{\mu\nu a} \widetilde{G}^a_{\mu\nu}\,,
\end{align}
as can be seen by substituting the equations of motion (\ref{wavetheta}) into the
divergence of (\ref{js2}). Notice that $\mathrm{U(1)_A}$ is still broken in the deconfinement \cite{Hegde:2011zg,Hegde:2011zg1}.

The total energy-momentum tensor is a sum of both electromagnetic $\Theta^{\mu\nu}_{\gamma}$ and gluonic ones $\Theta^{\mu\nu}_{\mathrm{g}}$
and the stress-energy tensor of the fluid $T^{\mu\nu}$.
The energy-momentum tensors of the free electromagnetic and gluonic fields are given by
\begin{align}
\Theta^{\mu\nu}_{\gamma}&=F^{\mu\lambda}F^{\nu}{}_\lambda - \frac{1}{4}g^{\mu\nu}F_{\alpha\beta}F^{\alpha\beta}\,,\\
\Theta^{\mu\nu}_{\mathrm{g}}&=G^{\mu\lambda a}G^{\nu}_\lambda{}^a - \frac{1}{4}g^{\mu\nu}G_{\alpha\beta}{}^a G^{\alpha\beta a}\,.
\end{align}
Their divergencies can be found by means of the equations of motion (\ref{mcs-eom}, \ref{grav-eom}),
\begin{align}
\partial_{\mu}\Theta^{\mu\nu}_{\gamma}&=-F^{\nu\lambda} j_{\lambda}- \C F^{\nu\lambda}\widetilde{F}_{\lambda\kappa}\partial^{\kappa}\theta \label{pTheta}\,,\\
\partial_{\mu}\Theta^{\mu\nu}_{\mathrm{g}}&=-g G^{\nu\lambda a} j_{\lambda}^a - \Cg G^{\nu\lambda a}\widetilde{G}_{\lambda\kappa}{}^a \partial^{\kappa}\theta \label{gTheta}\,.
\end{align}
Substituting this into the conservation law of the total energy-momentum,
\begin{align}
\partial_{\mu}(T^{\mu\nu}+\Theta^{\mu\nu}_\gamma+\Theta^{\mu\nu}_{\mathrm{g}})=0 \label{eq:fluidcons}\,,
\end{align}
we get (see also \cite{Ozonder:2010zy})
\begin{align}
\partial_{\mu}T^{\mu\nu}=F^{\nu\lambda} j_{\lambda} + g G^{\nu\lambda a} j_{\lambda}^a + \C F^{\nu\lambda}\widetilde{F}_{\lambda\kappa}\partial^{\kappa}\sm + \Cg G^{\nu\lambda a}\widetilde{G}_{\lambda\kappa}^a\partial^{\kappa}\sm
\,.\label{eq:mcs-hydro}
\end{align}
The first two terms on the right hand side are the standard terms for a
work done by external fields. The last two terms
is a work done by a Witten-like current \cite{GW}.
Finally, keeping that 
$\rho_5 \equiv j_5^0 = f^2 \partial^0 \sm$, we obtain an expression for the axial chemical potential, 
\begin{align}
\mu_5 = \ddd
\frac{\delta {\cal L}^{(2)}}{\delta\rho_5} = -\partial_0 \sm\,. \label{our_mu}
\end{align}
When boosted this turns into the Josephson-type equation $\mu_5 = - u^\mu \partial_\mu
\theta$, where $u^\mu$ will be chosen later as the velocity of the ``normal'' component, normalized by the condition $u_\mu u^\mu = -1$.
We also should assume, that $\partial_t \theta$ is slow varying in time, i.e. $\partial_\mu \theta \sim O(p^0)$, so the changes in chiral (axial) charge are small at the scale of QGP lifetime \cite{Andrianov:2012hq,Sadofyev:2010is}, otherwise the chiral chemical potential is not well-defined. Hydrodynamic and constitutive equations are then of order $O(p^2)$ and $O(p^1)$, respectively.

In summary, the system of hydrodynamic equations is given now by
\begin{align}
 \label{eq1enmomcon} \partial_{\mu}T^{\mu\nu} &= F^{\nu\lambda} (j_{\lambda} + \C \widetilde{F}_{\lambda\kappa} \partial^\kappa \sm) + G^{\nu\lambda a} (g j_{\lambda}^a + \Cg \widetilde{G}_{\lambda\kappa}{}^a \partial^\kappa \sm) \,,\\
 \label{eq1j5} \partial_\mu j_5^\mu &= -\frac{\C}{4}  F^{\mu\nu} \widetilde F_{\mu\nu} - \frac{\Cg}{4} G^{\mu\nu\,a} \widetilde{G}^a_{\mu\nu}\,,\\[5pt]
 \label{eq1j}  \partial_\mu j^\mu &= \partial_\mu j^{\mu a} = 0\,,\\[5pt]
  u^\mu\partial_\mu&\sm + \mu_5 = 0 \label{Josephson}\,,
\end{align}
where the last one is the Josephson-type equation. Corresponding constitutive relations in gradient expansion, satisfying (\ref{wavetheta}) and (\ref{eq1enmomcon}-\ref{eq1j5}), can be represented by
\begin{align}
 T^{\mu\nu} &= \left(\epsilon + P\right)u^\mu u^\nu + P g^{\mu\nu} + f^2
 \partial^\mu \theta\partial^\nu \theta+ \tau^{\mu\nu} \,, \label{enmom}\\
 j^\mu &= \rho u^\mu + \nu^\mu \,,\label{j}\\
 j^{\mu a} &= \rho^a u^\mu + \nu^{\mu a} \,,\label{ja}\\
 j_5^\mu &=  f^2 \partial^\mu \theta
  + \nu_5^\mu\label{js}\,,
\end{align}
where $\epsilon$, $P$, $\rho$, $\rho^a$ are the energy density, pressure, electric charge density and color charge density, respectively. Terms $\tau^{\mu\nu}$, $\nu^\mu$, $\nu^{\mu a}$ and $\nu_5^\mu$ denote higher-order gradient corrections and obey the Landau conditions
\begin{align}
\label{landau1}
u_\mu \tau^{\mu\nu} = 0\,, \qquad u_\mu \nu^{\mu} = 0\,, \qquad u_\mu \nu^{\mu a} = 0\,, \qquad u_\mu \nu_5^{\mu} = 0\,.
\end{align}
The stress-energy tensor $T^{\mu\nu}$ consists of two parts, an ordinary fluid component and a pseudoscalar ``superfluid'' component.
This modifies the Gibbs relation
\begin{align}
dP = s dT + \rho d \mu - f^2 d \left[\frac{1}{2} \partial^\mu\sm\,\partial_\mu\sm \right],\label{EOS}
\end{align}
where $s$ is the entropy density. Being additionally supported by the results of the Section~\ref{superfluidity}, we can describe the fluid content as a mixture of two components, originated from different parts of the fermionic spectrum,
\begin{itemize}
 \item[(1)] Zero and near-zero fermionic modes, which are involved into a potential (curl-free) motion of the chirality described by $j^5_\mu = f^2\,\partial_\mu\sm = \rho_5 u_{S \mu}$, where $u_{S \mu} \equiv \mu_5\,\partial_\mu \sm$ is the ``superfluid'' velocity.
 \item[(2)] The rest forming an ``axially-neutral'' component described by an electric $j_\mu = \rho u_\mu$ and color $j_\mu^a = \rho^a u_\mu$ currents and separated from the curl-free component by a finite gap.
\end{itemize}
There are numerous studies (e.g. \cite{Kharzeev:2007jp,Kharzeev:2001ev,Lappi:2006fp}) suggesting generation of a finite chirality in the processes of heavy ion collisions, so we can assume $\langle \rho_5 \rangle= f^2 \langle \mu_5 \rangle \neq 0$ within an event.
Such kind of initial conditions is not captured by lattice simulations ($\langle\rho_5\rangle=0$ after averaging over gauge configurations \cite{Buividovich:2009wi,Buividovich:2009wi1}), unless the finite $\langle\mu_5\rangle$ is introduced \textit{ad hoc} \cite{Yamamoto,Yamamoto1}.

As a final remark, one would naturally expect the energy-momentum tensor (\ref{enmom}) to be anisotropic at this order, since the electromagnetic fields are introduced. However, the anisotropy affects very weakly the anomalous effects in consideration: 
in \cite{Kalaydzhyan1, Kalaydzhyan2} it was shown that the anisotropy parameter enters the expressions for transport coefficients in a combination with chemical potentials, which are small in the case of heavy-ion collisions 
(this would not be the case, if we considered compact stars, see e.g. \cite{Ferrer:2010wz}).

\subsection{Phenomenological output, possible tests of the model\label{tests}}

Electric and magnetic fields in the fluid rest frame are defined as
\begin{align}
\label{EB}
 E^\mu = F^{\mu\nu}u_\nu, \qquad B^\mu = \tilde F^{\mu\nu}u_\nu \equiv  \frac{1}{2}\epsilon^{\mu\nu\alpha\beta}u_\nu F_{\alpha\beta}.
\end{align}
One can also rewrite these definitions in the following way
\begin{align}
F_{\mu\nu} =&\, \epsilon_{\mu\nu\alpha\beta}u^\alpha B^\beta + u_\mu E_\nu - u_\nu E_\mu,\\
\tilde F_{\mu\nu} =&\, \epsilon_{\mu\nu\alpha\beta}u^\alpha E^\beta + u_\mu B_\nu - u_\nu B_\mu\,,\label{Ftilde}
\end{align}
which we will use later.

The second term in first brackets (\ref{eq1enmomcon}) or equivalently the second term on r.h.s. of (\ref{mcs-eom}) has an interesting phenomenological interpretation as an additional electric current, induced by $\sm$-field, i.e. assosiated with the ``superfluid'' component. This current is conserved due to the Bianchi identity and vanishes in absence of either external electromagnetic fields or $\sm$. Let us split this term in three pieces using (\ref{Ftilde}):

\begin{align}
\label{phenom} j^{S}_\lambda &\equiv \C \widetilde{F}_{\lambda\kappa} \partial^\kappa \sm = \C \partial^\kappa \sm \left(
u_\kappa B_\lambda + \epsilon_{\kappa\lambda\alpha\beta}u^\alpha E^\beta - u_\lambda B_\kappa\right)\\
&= - \C \mu_5 B_\lambda + \C \epsilon_{\lambda\alpha\kappa\beta}u^\alpha \partial^\kappa \sm E^\beta - \C u_\lambda (\partial \sm \cdot B)\,.\nonumber
\end{align}
The first term in the sum is nothing but the Chiral Magnetic Effect (CME) \cite{Warringa}, i.e. generation of the electric current along the magnetic field.
The second term is analogous to the Chiral Electric Effect (CEE) \cite{CEE}, i.e. generation of the electric current perpendicular to applied
electric field and to both (normal and superfluid) four-velocities. The third term is a dynamical realization of the domain
wall polarization \cite{theta_domain,theta_domain1} or simply the fact that the ``would-be'' axions acquire an electric dipole moment in a magnetic field \cite{Wilczek:1987mv}.
Keeping in mind a wave-like profile of the $\sm$-field we could call this effect the ``Chiral Dipole Wave'' (CDW). First term contains $\mu_5$ which is in our case \textit{derived}
within QCD in contrary to other CME models. Last two terms are specific for the two-component fluid model and can be
considered as a possible experimental test of the model. A concrete quantitative prediction and comparison to the experimental data are beyond of the scope of this article and 
will be presented in a future publication (for some preliminary estimates, read \cite{Kalaydzhyan:2013gja}).

In order to study the CEE one should also take into account the response
of the medium (e.g. due to a finite electric conductivity), possible nonlinear effects (since
the field strength can be of the hadronic scale, if we take into account the Faraday's law), 
inhomogeneous character of the chemical potential, as well as the generation of an additional axial current due to the Chiral Electric Separation Effect \cite{Huang:2013iia}, etc.

The same analysis can be applied to the color current
\begin{align}
j^{S\, a}_\lambda \equiv \frac{\Cg}{g} \widetilde{G}_{\lambda\kappa}{}^a \partial^\kappa \sm\,,\label{phenom1}
\end{align}
which is the second term on the r.h.s. of (\ref{grav-eom}). The corresponding effects can be called ``Color CME'', ``Color CEE'' and ``Color CDW'', respectively. We are not focusing on these effects, since the color currents cannot be observed directly. It is also important to mention that both currents (\ref{phenom}, \ref{phenom1}) do not depend on the UV cutoff $\Lambda$ introduced in the preceding sections.

\subsection{Change in entropy and higher order gradient corrections}\label{appA}

In previous sections we kept higher-order corrections $\tau^{\mu\nu}$, $\nu^\mu$, $\nu^{\mu a}$, $\nu_5^\mu$ undetermined. 
These corrections incorporate possible dissipative effects (in presence of e.g. viscosity or electrical resistivity) 
and can be found from the constraint $\partial_\mu s^\mu \geq 0$ on the entropy 
current $s^\mu$ \cite{SS}. A priori it is not obvious whether they can interfere with the result (\ref{phenom}) or not.
Indeed, within the
ordinary hydrodynamics the chiral magnetic effect can be found as a part of the $\nu^\mu$ term \cite{Sadofyev:2010pr,Kalaydzhyan,Kalaydzhyan1}.
In this section we show that there are \textit{no} higher order corrections to the electric current $j_\mu$ arising in our case and affecting the 
phenomenological results of Section~\ref{tests}. 
Following \cite{Sadofyev:2010pr}, we transform the quantity
\begin{align}
 I \equiv u_\nu \partial_\mu T^{\mu \nu} + \mu \partial_\mu j^\mu + \mu_5 \partial_\mu j_5^\mu\,
\end{align}
using hydrodynamic and constitutive equations and equate both resulting expressions.

\subsubsection*{Constitutive equations}

\noindent Using the second law of thermodynamics $\epsilon + P = T s + \mu \rho$ and (\ref{enmom})-(\ref{js}), we can rewrite this quantity as
\begin{align}
I &= u_\nu \partial_\mu \left( \left(T s + \mu \rho \right)u^\mu \right) u^\nu + u_\nu \left(T s + \mu \rho\right) u^\mu \partial_\mu u^\nu + u_\nu \partial_\mu P g^{\mu\nu} \nonumber\\
& + u_\nu f^2 \partial^\nu\sm \,\partial_\mu \partial^\mu \sm + u_\nu f^2 \partial^\mu\sm \,\partial_\nu \partial^\mu \sm + u_\nu \partial_\mu\tau^{\mu\nu} \nonumber\\
& + \mu \partial_\mu \left(\rho u^\mu + \nu^\mu\right) + \mu_5 f^2 \partial_\mu\partial^\mu\sm + \mu_5 \partial_\mu \nu_5^\mu\,.
\end{align}

\noindent Using the normalization condition $u^\mu u_\mu = -1$, the Josephson equation $u^\mu\partial_\mu\sm + \mu_5 = 0$ and the following identities
\begin{align}
&\partial_\mu u^\mu = \mathrm{inv} = \partial_0 u^0_{\mathrm{(rest\,frame)}} = 0,\\
&u_\nu u^\mu \partial_\mu u^\nu = \frac{1}{2} \left( u_\nu u^\mu \partial_\mu u^\nu + u^\nu u^\mu \partial_\mu u_\nu\right) =u^\mu \partial_\mu (u^\nu u_\nu) \equiv 0,\\
&\partial_\mu (\rho u^\mu) = u^\mu \partial_\mu \rho + \rho \partial_\mu u^\mu = u^\mu \partial_\mu \rho = \mathrm{inv} = u^0 \partial_0 \rho = 0
\end{align}
we can simplify the expression for $I$ to
\begin{align}
I =& -T \partial_\mu \left(u^\mu s \right) - u^\mu \left\{ s \partial_\mu T + \rho \partial_\mu \mu  - f^2 \partial_\nu\sm \, \partial_\mu \partial^\nu \sm - \partial_\mu P\right\} + u_\nu \partial_\mu \tau^{\mu\nu}\nonumber\\
& + \mu \partial_\mu \nu^\mu + \mu_5 \partial_\mu \nu_5^\mu + 2\C \mu_5 E^\lambda B_\lambda - \frac{\Cg}{2} \mu_5 G^{\mu\nu a} \widetilde{G}_{\mu\nu}^a\,.
\end{align}
The sum in the curly brackets is identically zero due to the thermodynamic relation (\ref{EOS}). Also $u_\nu \partial_\mu \tau^{\mu\nu} = - \tau^{\mu\nu} \partial_\mu u_\nu$ because of the Landau frame condition (\ref{landau1}). This leads to the further simplification,
\begin{align}
\label{modif1}
 I = -T \left( \partial_\mu \left( s u^\mu\right) - \frac{\mu}{T}\partial_\mu \nu^\mu - \frac{\mu_5}{T}\partial_\mu \nu_5^\mu \right) -\tau^{\mu\nu} \partial_\mu u_\nu + 2C \mu_5 E^\lambda B_\lambda - \frac{\Cg}{2} \mu_5 G^{\mu\nu a} \widetilde{G}_{\mu\nu}^a\,.
\end{align}

\subsubsection*{Hydrodynamic equations}

\noindent Let us rewrite $I$ again using the hydrodynamic equations (\ref{eq1enmomcon}-\ref{eq1j5}). Then
\begin{align}
I =& F^{\nu\lambda} u_\nu \rho\, u_\lambda + u_\nu F^{\nu\lambda}\nu_\lambda + g G^{\nu\lambda a} u_\nu \rho^a\, u_\lambda + g u_\nu G^{\nu\lambda a}\nu_{\lambda}^a \nn\\
   & + \C u_\nu F^{\nu\lambda}\widetilde{F}_{\lambda\kappa}\partial^\kappa\sm + \Cg u_\nu G^{\nu\lambda a}\widetilde{G}_{\lambda\kappa}^a \partial^\kappa\sm\nn\\
   & + \C \mu_5 E^\lambda B_\lambda - \frac{\Cg}{4} \mu_5 G^{\mu\nu a} \widetilde{G}_{\mu\nu}^a \nn\\
 =& - E_\lambda\nu^\lambda + 2 \C \mu_5 E^\lambda B_\lambda - \frac{\Cg}{2} \mu_5 G^{\mu\nu a} \widetilde{G}_{\mu\nu}^a \,,
\end{align}
where we used the Josephson equation, the definitions of the electric and magnetic fields (\ref{EB}) and the corresponding inversed relations (\ref{Ftilde}).

\noindent Combining this result with (\ref{modif1}) we obtain
\begin{align}
 -T \left( \partial_\mu \left( s u^\mu\right) - \frac{\mu}{T}\partial_\mu \nu^\mu - \frac{\mu_5}{T}\partial_\mu \nu_5^\mu \right) = \tau^{\mu\nu} \partial_\mu u_\nu - E_\lambda\nu^\lambda\,.
\end{align}

\noindent Then dividing by $-T$ and adding $\ddd-\nu^\mu \partial_\mu \frac{\mu}{T} - \nu_5^\mu \partial_\mu \frac{\mu_5}{T}$ to the both sides we can rewrite the result as
\begin{align}
\label{entropycurrent}
 \partial_\mu \left( s u^\mu - \frac{\mu}{T}\nu^\mu - \frac{\mu_5}{T}\nu_5^\mu \right) = -\frac{1}{T}(\partial_\mu u_\nu) \tau^{\mu\nu} - \nu^\mu \left( \partial_\mu\frac{\mu}{T} - \frac{1}{T}E_\mu \right) - \nu_5^\mu \partial_\mu\frac{\mu_5}{T}\,.
\end{align}

Comparing this result with one of Son and Surowka \cite{SS}, we see that the term proportional to $\C$ is absent. This fact tells us that the divergence (\ref{entropycurrent}) is well defined, i.e. the entropy production is always nonnegative, and in contrary to \cite{SS} we do not need to add any additional terms to the entropy current.
Therefore, there are no leading-order corrections to $j^S_\lambda$ coming from the dissipative term $\nu^\mu$. So, all three anomalous effects are present already at the level of the hydrodynamic equations (\ref{eq1enmomcon}).

The expression for the entropy current remains unchanged also because the ``superfluid'' component itself has zero entropy.
Indeed, considering (\ref{entropycurrent}) in absence of dissipative corrections we obtain
\begin{align}
 \partial_\mu(s u^\mu) = 0,
\end{align}
i.e. only the ``normal'' component cotributes to the entropy current, which is a common property of a real superfluid. 
This fact would well agree with the long-range coherence of the chirality distribution \cite{Horvath,Horvath1,Horvath2,Horvath3,Horvath4,Horvath5,Ilgenfritz:2007xu,Ilgenfritz:2007b,Ahmad:2005dr,Thacker}, but 
should be studied more carefully, since the microscopic nature of $\sm$ is not known precisely.

\section{Conclusion}

In this paper we provided a novel treatment of the sQGP dynamics for the temperatures typical for recent and ongoing heavy-ion experiments. 
The main feature of our studies is a combination of nonperturbative methods applied to a general form of the theory, namely, the QCD coupled to QED. 
Our conclusion is that there can exsist a light (nearly massless) axion-like component of the quark-gluon plasma for the given range of temperatures. This 
component accommodates all the chiral and anomalous properties of the plasma and is responsible for a plenty of hypothetical effects leading to 
the \textit{local} $\mathcal{P}$- and $\mathcal{CP}$-violation in the strong interaction. The rest of the matter content of QGP obeys 
the hydrodynamic equations of a nearly ideal fluid. Both of the components together form some kind of a superfluid, which we call the ``chiral superfluid''.
This term should be understood in a relative sense as a two-component fluid with independent curl-free and normal motions. The separation between two motions is provided by
a gap in the Dirac spectrum observed in lattice QCD simulations with massless quarks. It is worth to mention, that in our case we do not obtain the 
light ``superfluid'' component as a Goldstone field of a broken continuous symmetry. Instead, it appears as a consequence of a nontrivial underlying vacuum structure.
At the same time, even not being a conventional superfluid, our system reproduces some usual properties, such as a (pseudo-)scalar nature of the 
``condensate'', zero entropy of the ``condensate'', Josephson-type relation, etc. Taking into account only regular configurations of the ``condensate'' 
we reserve the vortex-like solutions for the further studies. 
An important issue not covered in this article is the probe limit (quenching) for the lattice fermions.
It is not clear yet, whether the long-range structures survive in the full QCD or they are destroyed by dynamical fermions. 
As a crucial test of our model we propose some phenomenological effects (Section~\ref{tests}), which can be 
proven experimentally, i.e. a response of QGP to the presence of strong electromagnetic fields. A concrete description of the experimental consequences is currently
in progress.

\subsection*{Acknowledgments}

The author is grateful to Valentin I. Zakharov and Henri Verschelde for original ideas initiating
this study.
I also would like to thank Alexander Polyakov, Mikhail Shifman, Maxim Chernodub, Alexander Zhiboedov, Ingo Kirsch, Andreas Ringwald,
Mikhail Shaposhnikov, Dmitri Kharzeev, Jan Pawlowski, Edward Shuryak, Ismail Zahed,
Eliezer Rabinovici, Vitaly Bornyakov, Shiraz Minwalla, Guy Gur-Ari,
Piotr Surowka, Arata Yamamoto, Ivan Horv\'{a}th, Ilmar Gahramanov and Mikhail Isachenkov for interesting and useful
discussions, and Victor Braguta and Pavel Buividovich for helpful hints in doing numerics.
Numerical calculations have been performed at the GSI batch farm.


\end{document}